\documentclass[useAMS,usenatbib]{mnras}
\usepackage{graphicx}
\usepackage{bmpsize}
\usepackage{subfig}
\usepackage{hyperref}
\usepackage{amssymb}
\usepackage{amsmath}

 \def\be{\begin{equation}}
 \def\ee{\end{equation}}

\def\simless{\mathbin{\lower 3pt\hbox
   {$\rlap{\raise 5pt\hbox{$\char'074$}}\mathchar"7218$}}}   
\def\simgreat{\mathbin{\lower 3pt\hbox
   {$\rlap{\raise 5pt\hbox{$\char'076$}}\mathchar"7218$}}}   
\def\etal{{\rm et al.}}

\def\solm{{\rm M}_\odot}

\def\apj{{ApJ}}
\def\apjl{{ApJL}}

\def\mnras{{MNRAS}}

\title[A clustered origin for isolated massive stars] {A clustered origin for isolated massive stars}
\author[Lucas \etal]{William E. Lucas$^1$\thanks{E-mail: wel2@st-andrews.ac.uk}, Matus Rybak$^{1,2}$, Ian A. Bonnell$^1$, and Mark Gieles$^3$\\
\normalsize{$^{1}$ Scottish Universities Physics Alliance (SUPA), School of Physics and Astronomy, University of  St Andrews, } \\
\normalsize{North Haugh, St Andrews, Fife KY16 9SS, UK}\\
\normalsize{$^{2}$ Leiden Observatory, Leiden University, PO Box 9513, NL-2300 RA Leiden, the Netherlands}\\
\normalsize{$^{3}$ Department of Physics, University of Surrey, Guildford GU2 7XH, UK}\\}

\date{\today}

\begin{document}

\date{Accepted 00 June 0000. Received 0000 December 00; in original form 15 June 1215}

\pagerange{\pageref{firstpage}--\pageref{lastpage}} \pubyear{2009}

\maketitle

\label{firstpage}

\begin{abstract}
High-mass stars are commonly found in stellar clusters promoting the idea that their formation occurs due to 
the physical processes linked with a young stellar cluster. It has recently been reported that isolated high-mass stars are present in the Large Magellanic Cloud. Due to their low velocities it has been argued that these are high-mass stars which formed without a surrounding stellar cluster. In this paper we present an alternative explanation for the origin of these stars in which they formed in a cluster environment but are subsequently dispersed into the field as their natal cluster is tidally disrupted in a  merger with a higher-mass cluster. They escape the merged cluster with relatively low velocities typical of the cluster interaction and thus of the larger scale velocity dispersion, similarly to the observed stars. $N$-body simulations of cluster mergers predict a sizeable population of low velocity ($\le 20\,\mathrm{km}\,\mathrm{s}^{-1}$), high-mass stars at distances of $>20$ pc from the cluster. High-mass clusters in which gas poor mergers are frequent would be expected to commonly have halos of
young stars, including high-mass stars, that were actually formed in a cluster environment. 
\end{abstract}


\section{Introduction}

Although relatively low in number, massive stars play a key role in the evolution of the Universe. They have a huge impact on both star and planetary formation processes (\citealt{Hollenbachetal1994}; \citealt{ZinneckerYorke2007}; \citealt{Deharvengetal2010}) and on the evolution of galaxies (\citealt{Vossetal2010}; \citealt*{Hopkinsetal2012}; \citealt{Walchetal2015}; \citealt{Girichidisetal2016}). Unfortunately, our knowledge of how high-mass stars form is relatively limited.  They are relatively rare and hence are located at large distances. Their early evolutionary stages are usually obscured by surrounding dust and they are predominantly found in dense stellar clusters. 

In spite of all these challenges, two major theories of high-mass star formation have been developed. In the monolithic collapse model (\citealt*{KruMcKKle2005}; \citealt{KruMcK2008}), the mass of a protostar is set by the mass of a physically distinct core in a molecular cloud. In such a model, a massive star could be formed from a small molecular cloud far away from any nearby star cluster, although it would likely need to have significant magnetic support to increase the Jeans mass and hence limit fragmentation \citep{HenTey2008}. The alternative model is based on competitive accretion (\citealt{Bonnelletal2001a}; \citealt*{Bonnelletal2004}) in a forming stellar cluster whereby the overall gravitational potential helps funnel gas down to the centre, enabling stars located there to accrete at higher rates and from regions well beyond their natal core. In such a scenario, high-mass stars always form in clustered environments. 

Although observations do show that high-mass stars are most commonly found in clustered environments, with Galactic estimates of at most a few percent being consistent with formation in isolation \citep{deWitetal2004,deWitetal2005}, there have been recent reports of isolated high-mass stars on the outskirts of massive clusters such as 30 Doradus \citep{Bressertetal2012}. Bressert et al. found $\approx 15$ isolated O-star candidates, located at distances from 14 to 130 pc from R136, out of a sample of 800 O-stars. This study considered isolated stars to be those with no surrounding clusters. They also excluded high velocity runaways, defined as having line-of-sight velocities more than one $\sigma = 10.50 \,\mathrm{km}\,\mathrm{s}^{-1}$ from the O-star mean of $270.73 \,\mathrm{km}\,\mathrm{s}^{-1}$, as these could be described as escapees from encounters in the cluster. Recently, \citet{Stephensetal2017} selected seven seemingly isolated massive young stellar objects (MYSOs) in the Large Magellanic Cloud and found that all are actually surrounded by previously undetected clusters on order a few hundred low-mass stars. This would lend weight to the idea that massive stars will always form in a clustered environment, but then demands a solution as to how they later enter the truly isolated, non-clustered state of the \citet{Bressertetal2012} stars.

The simulations of \citet*{BanKroOh2012} were able to produce slow runaways which may resemble isolated stars from binary scatterings in a single massive cluster. Only $\sim 10$ per cent of O-stars in 30 Doradus reside in the handful of dense clusters \citep{Doranetal2013}, and with available line-of-sight velocities we are not able to determine whether each O-star can be traced back to an origin in a dense cluster. The large fraction of O-stars in relative isolation still lacks an explanation. In this paper we consider whether a cluster merger can also explain these stars, ejecting them at low velocities and leaving them in isolation. In Section~\ref{s:hier_clust_form} we review cluster formation models and how these can disperse a population of high-mass stars into the field. In Sections~\ref{s:num_sims} and \ref{s:mergers} we present the setup of numerical simulation of this scenario and then discuss the results.

\section{Hierarchical cluster formation}\label{s:hier_clust_form}

The formation of stellar clusters involves the fragmentation of a molecular cloud into hundreds to many thousands of self-gravitating objects. Fragmentation  requires that local collapse is faster than the collapse of the cloud as a whole. This occurs most easily when the dimensionality of an object is reduced such that a spherical object becomes sheetlike or preferably filamentary. The free-fall timescale of a small region within a cloud can be compared to the timescale for the cloud as a whole to determine whether a local or global collapse will occur. For a filament, the local density within the subregion is higher and thus the free-fall timescale shorter than that for the cloud as a whole, which depends on the mean density interior to the subregion, making fragmentation the likely outcome \citep{Bonnell1999}.

Turbulence in molecular clouds naturally produces filamentary structures observed in the infrared \citep{Molinarietal2010} which are ideal sites for fragmentation. Numerical simulations show that stellar clusters form from a hierarchical  process starting from fragments infalling along filaments, and growing with the merging of small groups and subclusters that infall into the larger scale gravitational potential (\citealt{BonBatVin2003}; \citealt{Maschbergeretal2010}). The presence of gas dissipates excess kinetic energy, helping to ensure a dense, bound cluster results.  

For more massive clusters ($\gtrsim 10^4 \solm$), this process is likely to continue over longer time periods of several Myr as the merging clusters are assembled from components spread over larger distances (Smilgys \& Bonnell 2017, submitted). The longer timescale to build up these clusters implies the gas is likely to be either accreted or dispersed before the merger process is completed. In addition, they will interact with larger impact velocities due to the larger gravitational potentials as the  clusters grow in mass.  These ``dry'' mergers will be unable to  dissipate the excess kinetic energy making it harder to  accrete the merging cluster in its entirety. Instead, the incoming cluster is likely to be tidally disrupted, spilling some of its stars into the surrounding field. It is this phase we wish to explore.

Recent observational studies of the cluster R136 in the heart of 30 Doradus shows evidence of complex dynamics \citep{Henault-Brunetetal2012} and a potential hierarchical merger \citep{FujiiZwart2011}. This is further suggested by \citet{Sabbietal2012} who found two distinct stellar populations in the core of 30 Doradus, with the North-East clump located 4.5 pc away from the centre of R136 and with a relative velocity of 10 to $15\,\mathrm{km}\,\mathrm{s}^{-1}$.

We investigate whether the massive stars, which formed in the core of a relatively small cluster, can be dispersed into the field during the course of a two-cluster merging event with a more massive counterpart. In such a scenario, the stars will be travelling at relatively low velocities typical of the cluster interaction, yet can still travel the required 30 pc in a few Myrs. Their low velocities would ensure they are not classified as runaways, yet they would be in isolation. This mechanism could provide an alternative explanation for the high-mass stars seen by \citet{Bressertetal2012} without a need to invoke isolated high-mass star formation.

\subsection{Tidal disruption of merging clusters}\label{ss:tidal_disruption}

In order for the process outlined above to be effective, the incoming cluster must be tidally disrupted in order to spill the high-mass stars into the field.  The clusters are virialised systems such that they only need an increase of a factor of 2 in their internal kinetic energies in order to become unbound. The specific kinetic energy imparted into the secondary cluster, of size $r_2$, as it approaches within impact parameter $b$ of the primary cluster can be estimated as
\begin{equation}
\Delta E_{\rm tidal} \sim \left(\frac{\partial \Phi}{\partial r}\right)_b r_2.
\end{equation}
If this exceeds the kinetic energy already contained in the cluster, $E_{\rm kin} \approx 1/2 \frac{GM_2}{r_2}$,
then the cluster will be disrupted, allowing its constituent stars to spread into the field. Approximating the cluster potential as $\Phi = -GM_1/r$, this is equivalent to saying that the in falling cluster must penetrate within
\begin{equation}
r_\mathrm{min} < q^{-1/2} r_2, 
\end{equation}
where $q=M_2/M_1$ is the mass ratio of the clusters. If the clusters have equal densities, then this can be rewritten as
\begin{equation}
r_\mathrm{min} < q^{-1/6} r_1.
\end{equation}
This effectively means that if the secondary cluster penetrates within 1 to 1.4 times the effective radius of the primary cluster (for $q$ between $0.1$ and $1$), it should be tidally disrupted.

\citet{Spitzer1958} described this in more detail, though it should be noted that these derivations adopt a simple model with point-mass clusters. Relations which treat them fully as extended mass distributions are given by \citet*{GneHerOst1999} and \citet{Gielesetal2006}. The energy gained by the secondary cluster during an encounter is, relabelling some terms for our purpose here,
\begin{equation}
	\Delta U = \frac{1}{2} M_2 \left ( \frac{2GM_1}{r_\mathrm{min}^2 v_\mathrm{max}} \right )^2 \frac{2}{3}\bar{r_2^2}
\end{equation}
where the subscripts 1 and 2 indicate the primary and secondary clusters, $M$ is mass, $\bar{r^2}$ the cluster mean square radius, $r_\mathrm{min}$ the closest approach of the clusters and $v_\mathrm{max}$ the maximum velocity. For a virialized cluster with total energy equal to half the gravitational binding energy, we find the criterion for disruption in a single encounter
\begin{equation} \label{eq:disruption_condition}
	r_\mathrm{min}^2 v_\mathrm{max} < \sqrt{\frac{2G}{\pi\gamma}} \frac{M_1}{\sqrt{\rho_2}},
\end{equation}
where $\gamma$ is the constant derived from the integration of total gravitational energy and is e.g. $0.465$ for a uniform density sphere. Using this value of $\gamma$ , the criterion may be recast as
\begin{equation}
	\left( \frac{r_\mathrm{min}}{3\,\mathrm{pc}} \right)^2 \left( \frac{v_\mathrm{max}}{10\,\mathrm{km}\,\mathrm{s}^{-1}} \right) < 2.70 \left( \frac{M_1}{10^5 \solm} \right) \left( \frac{10^3 \solm \,\mathrm{pc}^{-3}}{\rho_2} \right)^\frac{1}{2}.
\end{equation}

This form of the criterion exposes the preferred collision parameters for disruption. The left hand side of the inequality makes it clear that it will be close and slow ($\sim 1\,\mathrm{pc}$, $\sim 10 \,\mathrm{km}\,\mathrm{s}^{-1}$) encounters which lead to disruption of the secondary cluster, as needed for mergers. The right hand side in turn favours more massive primaries and less dense secondaries. It may be expected that a higher cluster density indicates a higher mass; in this case, disruption is easier for lower mass ratios $q$. If we use \citet*{DavClaFre2010} to estimate cluster properties at the end of contraction that takes place during their formative period (before the expansion described by \citealt{ClaBon2008}), the cluster density-mass relation can be estimated to be $\rho \sim M^{5/3}$, or a mass-radius relation of $r \sim M^{-2/9}$. This makes the relation on the right hand side of Equation~\eqref{eq:disruption_condition} $M_1/M_2^{5/6}$ -- very nearly $1/q$ -- indicating that disruption is easier for smaller $q$.

Equation~\eqref{eq:disruption_condition} may be rewritten further. If we take the velocity of the secondary cluster to be zero at infinity and treat the primary to be a point mass, then $v_\mathrm{max}^2 = 2GM_1/r_\mathrm{min}$. Substituting in both this and the density of the secondary cluster $\rho_2 = 3M_2 / 4\pi r_2^3$, we find that the criterion for disruption has a more traditional form resembling the Roche radius (e.g. \citealt{BinneyAndTremaine} Ch. 8.3)
\begin{equation}
	r_\mathrm{min} < r_2 \left( \frac{4}{3\gamma} \frac{M_1}{M_2} \right)^\frac{1}{3}
\end{equation}
which is effectively a comparison of mass densities at the point of pericentre passage.

The three forms of the criterion above point to tidal disruption taking place for close and slow encounters with a small mass ratio $q$ and lower densities in the secondary cluster. Once the encounter has taken place, the stars in the disrupted cluster then have the kinetic energy of the cluster's internal motions plus the kinetic energy of the infalling cluster, and hence can be unbound from the primary cluster. A significant proportion of these stars can then escape the system. 

\begin{figure*}
\begin{center}
\includegraphics[width=\textwidth]{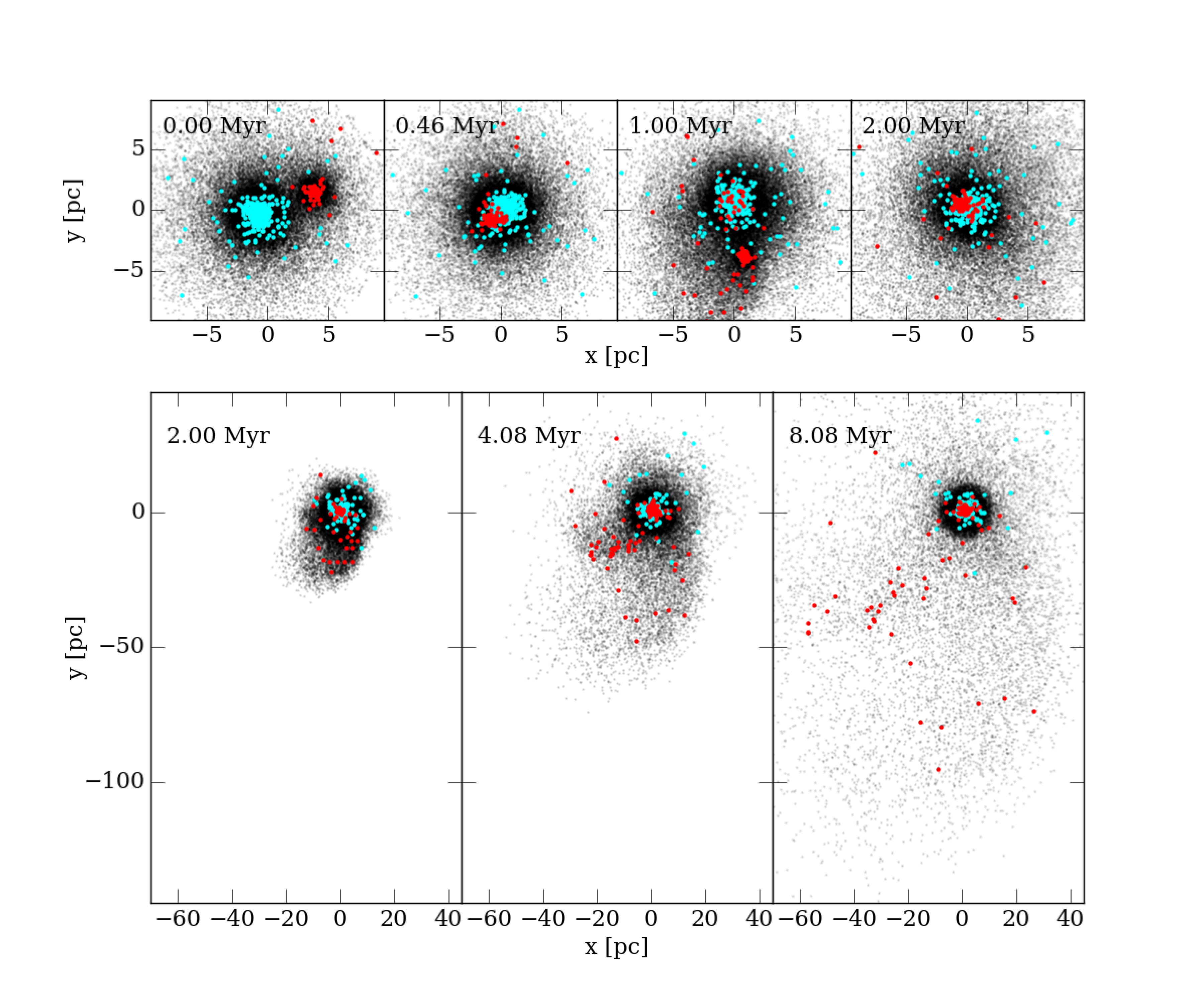}
\caption{\label{fig:clustdisp} The evolution of the merging clusters is shown for the fully resolved simulation with $\epsilon = 0.333, q = 0.2, b = 1.8$ pc, v$_{\rm rel}$ = $10.0\,\mathrm{km}\,\mathrm{s}^{-1}$, with half the stars being mass segregated and the others being on random orbits. The simulation includes $10^5$ particles, and was run on a tree-code using gravitational softening with 10,000 AU. Massive stars from the smaller (red) and main clusters (blue) have been highlighted while stars which have died as supernovae are not shown. The upper panel shows snapshots of the inner 10x10 pc region at times of  $0, 0.46, 1.00$ and $2.00$ Myr, while the bottom panel shows a zoomed out view at times of $2.0, 4.08$ and $8.08$ Myr showing the large scale effect of the tidal disruption of the lower-mass cluster. The high-mass stars in the smaller cluster are mostly dispersed into the field at velocities of order the interaction velocity of $10\,\mathrm{km}\,\mathrm{s}^{-1}$. }
\end{center}
\end{figure*}

\section{Numerical simulations}\label{s:num_sims}

An initial exploration of the parameter space was performed using a total of 75 simulations of cluster mergers, followed for 4 Myr, using the \textsc{nbody6} package \citep{Aarseth1999} with $10^4$ particles. Following this, full simulations with $10^5$ particles were followed for 8 Myr using a tree-code gravity solver from \textsc{sphNG} \citep*{BatBonPri1995}. This used gravitational softening to remove the effects of binary encounters, and it is these simulations on which we principally focus.

Initial conditions were set using the \textsc{mameclot}\footnote{https://github.com/mgieles/mameclot} package. For each cluster, the stellar positions were randomly drawn from an isochrone model which has a potential \citep{Henon1959}
\begin{equation}
\phi(r)=-\frac{GM}{r_{s}+a(r)}
\end{equation}
where \emph{M} is the total mass of the cluster and $r_{s}$ is the scale radius, and where $a(r)=\sqrt{r_{s}^{2}+r^{2}}$. The corresponding density profile can be found by applying Poisson's equation to yield:
\begin{equation}
\rho(r)=M\cdot\frac{3(r_{s}+a)a^{2}-r^{2}(r_{s}+3a)}{4\pi(r_{s}+a)^{3}a^{3}}
\end{equation}
with a scale radius of $r_s=1$ pc for the primary cluster. The half-mass radius for such a system is $r_{h} \approx 3.03 r_s$. Velocities are drawn from the distribution function \citep{Henon1960}.

For the initial exploration, the two clusters were set up to interact with impact parameters between 0 and 4pc and relative velocities between 6 and $20\,\mathrm{km}\,\mathrm{s}^{-1}$. The mass ratio of the two clusters varied from $q=0.1$ to $q=0.3$ and the secondary's radius was set from the concentration parameter $\epsilon$ where $r_{\mathrm{h},2}/r_{\mathrm{h},1} = q^{\epsilon}$. The simulations spanned concentration parameters of $\epsilon =0.167$ to $\epsilon=0.5$. An $\epsilon =1/3$ implies equal densities while $\epsilon>1/3$ implies a satellite with a higher mass density and vice versa.

The \textsc{nbody6} simulations were run at $1/10$ resolution  ($N = 10,000$) particles, such that each particle represents 10 stars of a given mass, while preserving the shape of the \citet{Kroupa2001} initial mass function (IMF). The IMF itself ran from $0.1$ to $100\solm$. This allowed a greater flexibility computationally with the only cost that of a small increase in two-body effects. The large-scale tidal interactions that we are investigating were not affected. Mass segregation was approximated by rescaling the positions of the higher-mass stars $M>5\solm$ by a factor $1/5$. For $q$ of $0.3$, giving secondary clusters that were more bound, and the highest impact parameters, disruption did not always take place during the first pass through pericentre.

As the simulations were intended to examine the process described in the previous section, the clusters were simulated in isolation with no application of an external tidal field. It is left to further investigation to determine the role (if it exists) played by the still poorly understood internal tidal structure of the LMC.

The tree-code gravity simulations used the same initial conditions as the initial runs, but were fully resolved such that particle masses were appropriate for individual stars. Mass segregation was implemented using the method of \citet*{BauDeMKro2008} which inversely sorts the stars' masses against their specific orbital energies. This was applied at three levels by sorting none of the stars, half of them, or all of them. We also used three mass ratios of $q = 0.1$, $0.2$ and $0.3$, for a total of nine variations on the initial conditions. They all used an impact parameter of $b = 1.8\,\mathrm{pc}$ and relative speed of $v_\mathrm{rel} = 10\,\mathrm{km}\,\mathrm{s}^{-1}$, values indicated by the \textsc{nbody6} grid to demonstrate the disruption-merger event.

As mentioned above, the tree-code used gravitational softening to remove binary systems and any ensuing ejections due to interactions with them. We tested two softening lengths of $10^3$ and $10^4\,\mathrm{AU}$ in our initial tests of the $q = 0.2$, non-mass segregated setup. With the smaller softening length, $110$ isolated stars were ejected by the end of the test at 4 Myr, while the larger value gave $108$ stars. We took this to indicate that by this stage the stars were being ejected only by the large scale streaming motions during the violent relaxation of the two clusters, with two-body relaxation playing almost no role. (How stars were determined to be isolated is described in Section~\ref{ss:clust_origin}.) The remainder of the simulations were performed using $10^4\,\mathrm{AU}$ alone. Though this precluded the formation of the slow runaways found by \citet{BanKroOh2012}, another potential avenue for the formation of isolated high-mass stars, our suppression of binary ejections ensures that we could not confuse these stars with those escaping due to the cluster merger and tidal disruption, the aim of this paper.

\begin{table*} \centering
	\caption{Tree-code simulations using a softening length of $10^4\,\mathrm{AU}$ with clusters colliding with $b = 1.8\,\mathrm{pc}$ and $v_\mathrm{rel} = 10\,\mathrm{km}\,\mathrm{s}^{-1}$. Given here are the number of isolated massive stars at three points during the simulations (2.00, 4.08 and 8.08 Myr), as well as their mean and maximum mass at the $8.08\,\mathrm{Myr}$ point. A star was classified as massive and isolated if it was above $5 \solm$ and had a stellar number density below $1\,\mathrm{pc}^{-3}$. Stars were also excluded if the simulation time exceeded the stellar lifetime obtained by interpolating the sum of the hydrogen and helium burning timescales of \citet{MeyMae2003} for stars initially rotating at $300\,\mathrm{km}\,\mathrm{s}^{-1}$. In two cases the number of isolated stars fell between $4$ and $8\,\mathrm{Myr}$. Very few isolated stars were lost as supernovae (and none at all for the fully-segregated runs), so this was due instead to stars on large but bound orbits being found in low density space earlier on. Then at the later time, having returned to a more clustered environment, they no longer met the criterion for being isolated.}
	\begin{tabular}{@{}lcccccccc}
		\hline
	Mass ratio $q$ & Mass segregation & $N_\mathrm{isolated}(2\,\mathrm{Myr})$ & $N_\mathrm{isolated}(4\,\mathrm{Myr})$ & $N_\mathrm{isolated}(8\,\mathrm{Myr})$ & $\langle m_\mathrm{isolated}\rangle/\solm$ & $m_\mathrm{isolated,max}/\solm$ \\
		\hline
		0.1 & 0.0 & 50 & 66 & 77 & 9.97 & 28.8 \\
		0.2 & 0.0 & 70 & 112 & 100 & 9.92 & 26.6 \\
		0.3 & 0.0 & 71 & 130 & 119 & 9.49 & 23.4 \\
		0.1 & 0.5 & 14 & 32 & 40 & 10.2 & 24.7 \\
		0.2 & 0.5 & 28 & 60 & 62 & 8.66 & 19.5 \\
		0.3 & 0.5 & 36 & 71 & 74 & 8.55 & 24.0 \\
		0.1 & 1.0 & 0 & 0 & 15 & 6.48 & 9.06 \\
		0.2 & 1.0 & 0 & 14 & 15 & 7.19 & 12.5 \\
		0.3 & 1.0 & 1 & 18 & 16 & 9.13 & 25.5 \\
		\hline
	\end{tabular}
	\label{tab:sim_names}
\end{table*}


\begin{figure}
\vspace{-0.4cm}
\begin{center}
\subfloat[$q=0.2$, no mass segregation.]{
\includegraphics[width=0.47\textwidth]{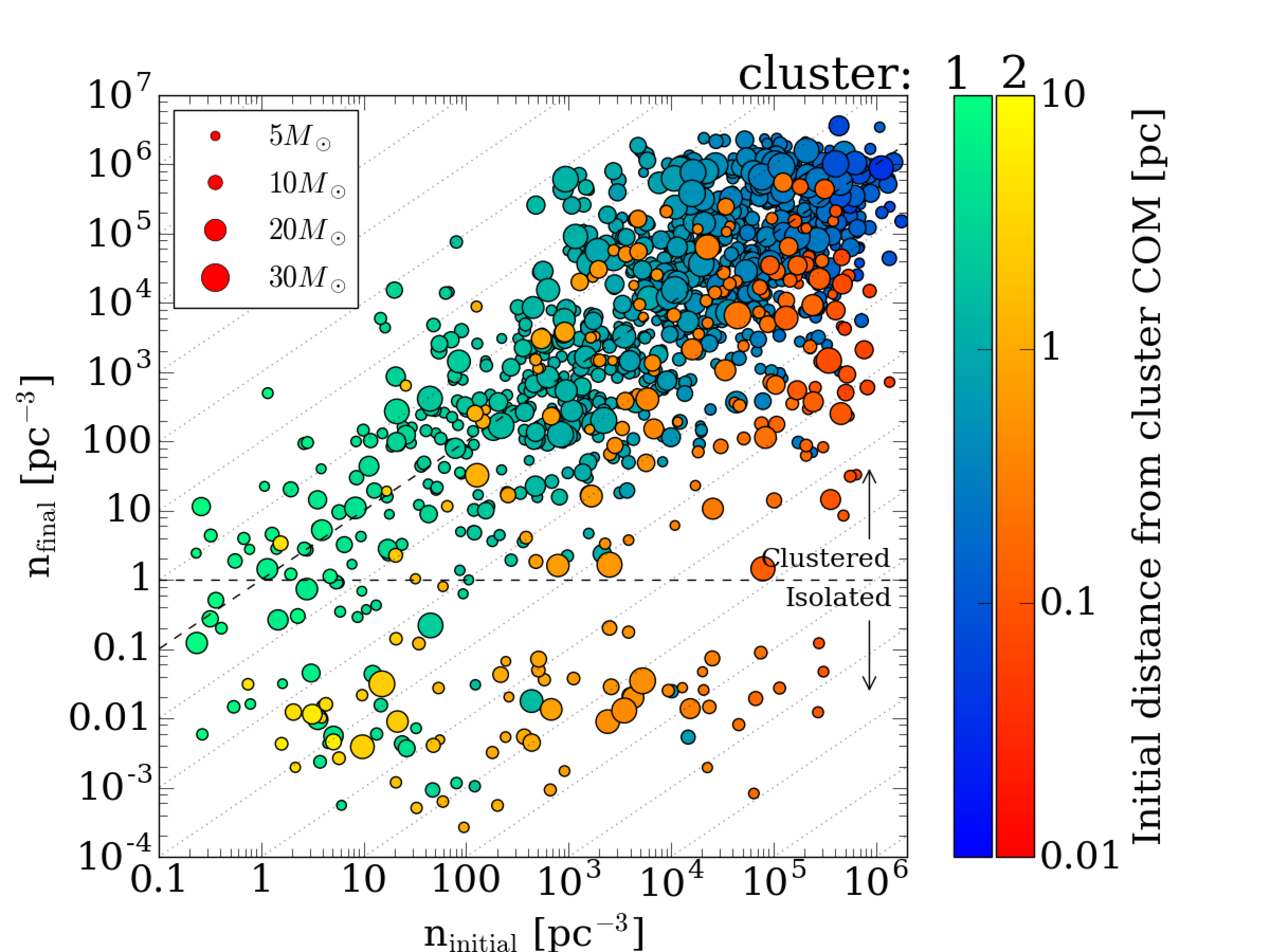}
}\\
\vspace{-0.2cm}
\subfloat[$q=0.2$, partial mass segregation.]{
\includegraphics[width=0.47\textwidth]{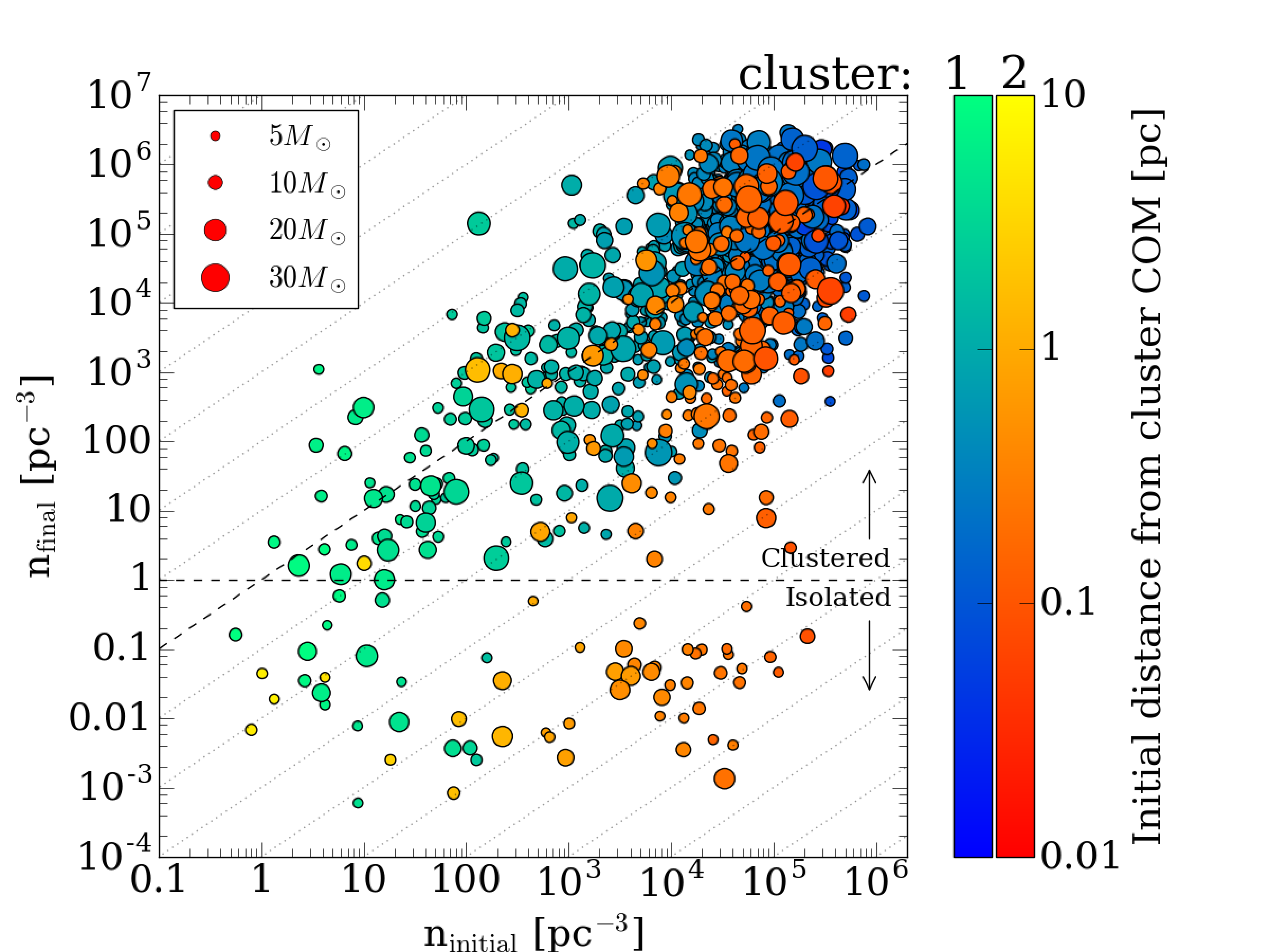}
}\\
\vspace{-0.2cm}
\subfloat[$q=0.2$, full mass segregation.]{
\includegraphics[width=0.47\textwidth]{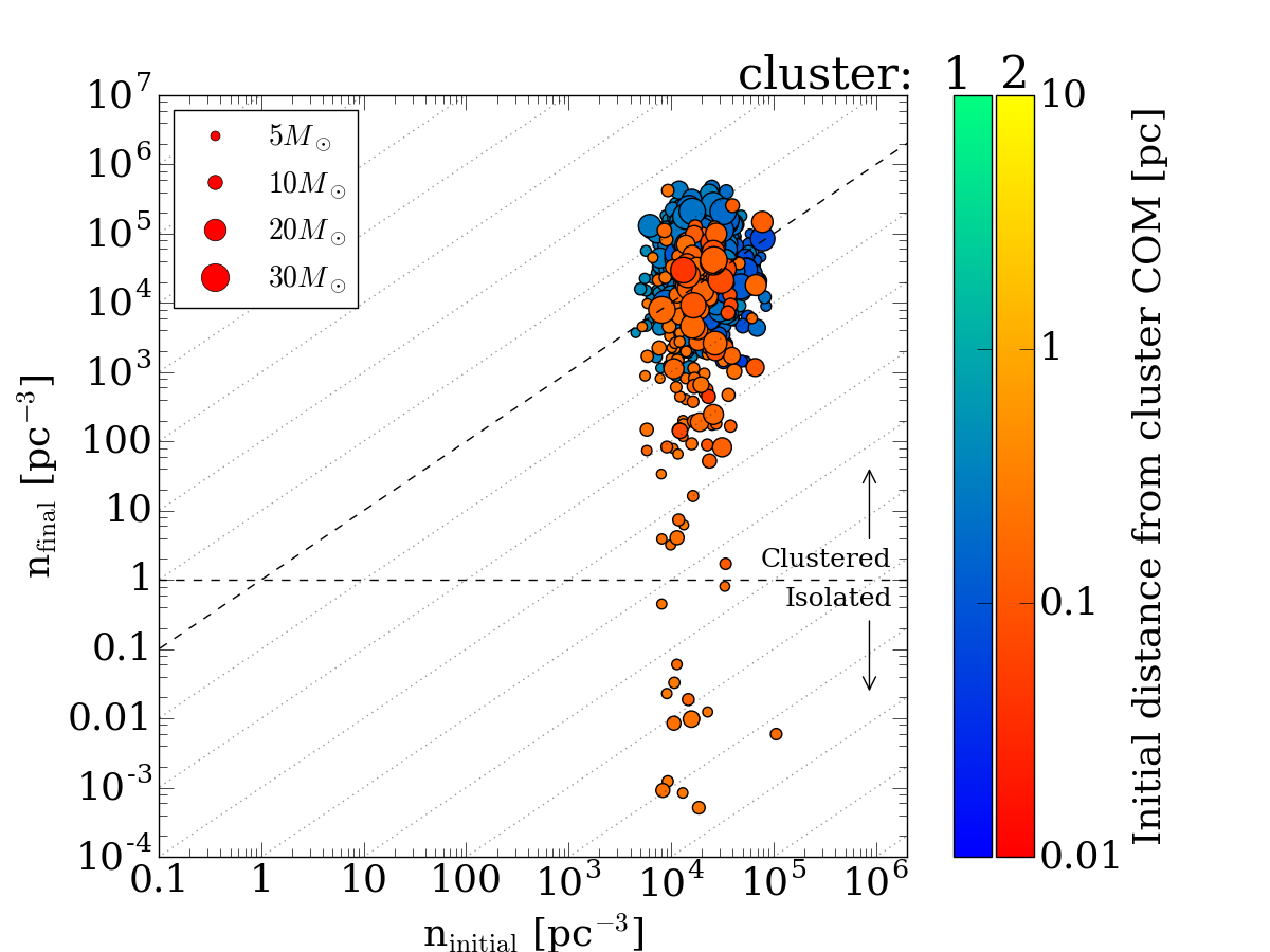}
}\\
\caption{Initial and final (8.08 Myr) number densities for the high-mass stars in both the primary (green/blue) and secondary (red/yellow) clusters using tree-code runs. The colours themselves indicate the initial distance of each star from its original cluster's centre of mass. The diagonal dashed lines show where the densities were unchanged, and the dotted lines mark changes by factors of ten; the vertical dashed lines show the point beyond which we considered a star to be isolated. Stars which would have died by the final time are not shown. With increasing mass segregation the two clusters contributed fewer stars to the isolated population until the primary finally contributed none at all. The average mass of liberated stars fell in the same way. In all cases here some stars' number densities dropped by factors of $\sim 10^7$.}
\label{fig:ninit_nfinal}
\end{center}
\end{figure}

\section{Cluster mergers} \label{s:mergers}

As expected from \S 2.1 and the initial parameter space exploration, the clusters as simulated by our tree-code undergo a tidal disruption and merger while throwing off stars with the excess energy. Figure~\ref{fig:clustdisp} shows the evolution of a cluster merger event for a mass ratio of $q=0.2$ and an impact parameter of $1.8$ pc, relative velocity of $10\,\mathrm{km}\,\mathrm{s}^{-1}$ and partial mass segregation, as performed with the tree-code. The secondary penetrates within the half-mass radius of the primary cluster and is effectively disrupted in the interaction. Figure~\ref{fig:clustdisp} highlights the massive stars $m \ge 5 \solm$,  in both the primary (blue) and secondary (red) clusters.  

The dispersed secondary forms into a tidal tail allowing most of the stars to become bound to the primary cluster while the rest escape with the excess kinetic energy. This holds for the high-mass stars as well as for the lower-mass stars, allowing a population of high-mass stars to escape the dense clusters and spread into the surrounding field. The bottom three panels of Figure~\ref{fig:clustdisp} show the formation of the tidal tails and the population of high-mass stars therein. These high-mass stars are given excess velocities of order the interaction velocity of some 10 to $20\,\mathrm{km}\,\mathrm{s}^{-1}$ and can thus reach distances of several tens of parsecs within a few Myr of the encounter.

Hitherto unmentioned is the fact that by the simulations' end time of 8.08 Myr we would expect many of the highest mass stars present to have already died. Our tree-code did not itself account for stellar evolution, so we determined the lifetimes of our massive stars by interpolating their masses with the summed hydrogen and helium burning timescales of \citet{MeyMae2003} for stars with initial rotational velocities of $300\,\mathrm{km}\,\mathrm{s}^{-1}$. Typical initial rotational velocities range from a few tens of to several hundred kilometres per second (e.g. \citealt{WolStrDro2006}; \citealt{Markovaetal2014}), though the lifetimes for the most massive stars depend only slightly on this quantity. In results we give for specified times in the simulations, we exclude stars with shorter lifetimes. This is equivalent to stating that the all the stars were formed instantaneously at the beginning of the simulation. Although this is unrealistic, it prevents the introduction of another free parameter.

\subsection{A clustered origin of ``isolated'' high-mass stars} \label{ss:clust_origin}

\begin{figure}
\begin{center}
\includegraphics[width=0.52\textwidth]{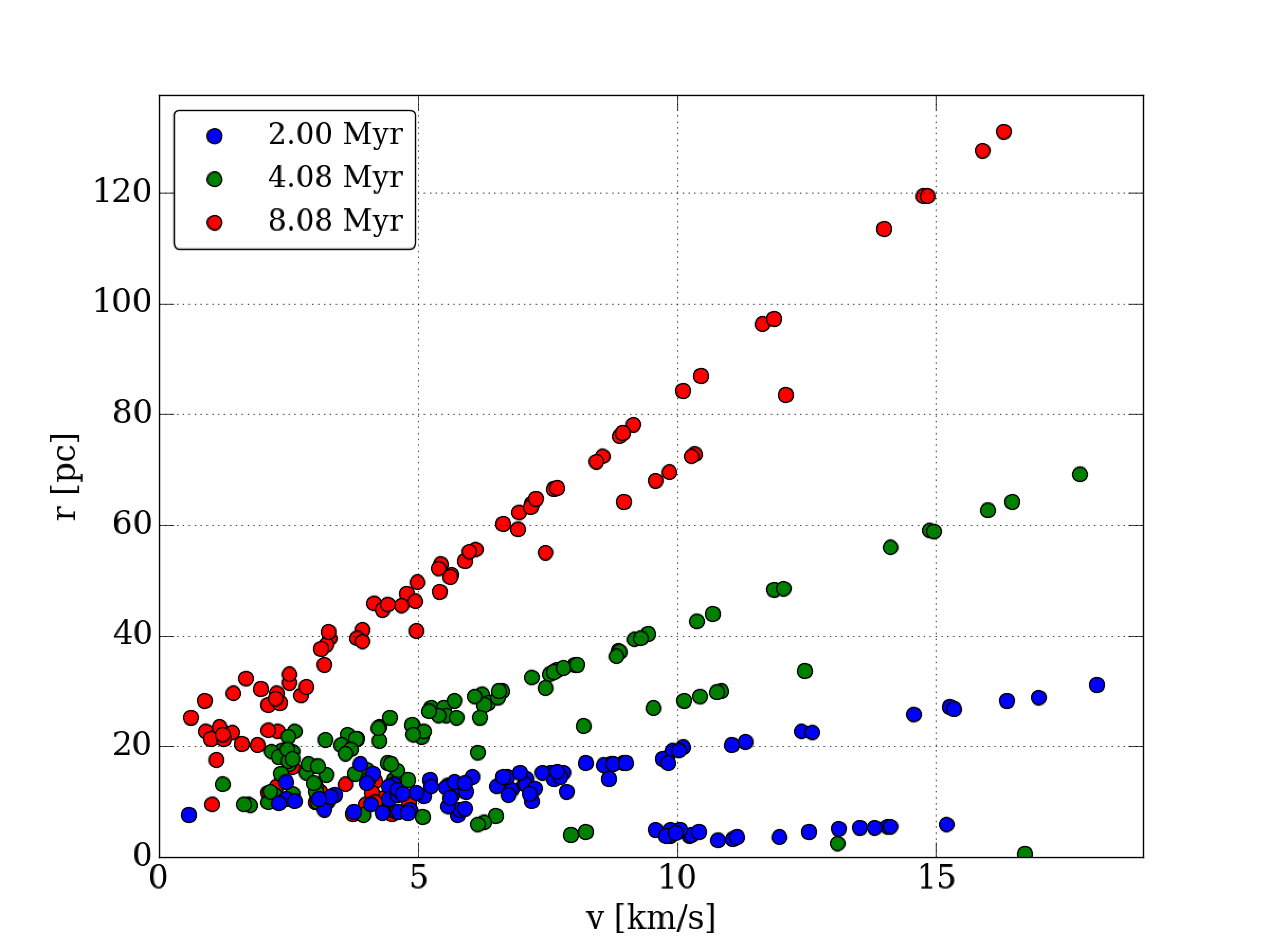}
\caption{Velocity-position diagram of high-mass isolated stars as determined at the final timestep in the fully-resolved $q=0.2$ non-mass segregated simulation. Shown are the stars which would still be alive at $2.00$, $4.08$ and $8.08$ Myr. The stars were thrown off from the encounter at velocities between a few and $\lesssim 20\,\mathrm{km}\,\mathrm{s}^{-1}$. At each time a clear marker of the second disruption event can be seen, extending to lower maximum velocities of $\lesssim 15\,\mathrm{km}\,\mathrm{s}^{-1}$. These velocities are in excess of the local escape speed and by the final time the stars have reached distances of 20 to between 80 and 140 pc. The velocities are not dissimilar to those of the vast majority of the bound stars remaining in the cluster and thus would not be identified as high-mass runaways.}
\label{fig:posvel}
\end{center}
\end{figure}

\begin{figure*}
\begin{center}
\includegraphics[width=0.9\textwidth]{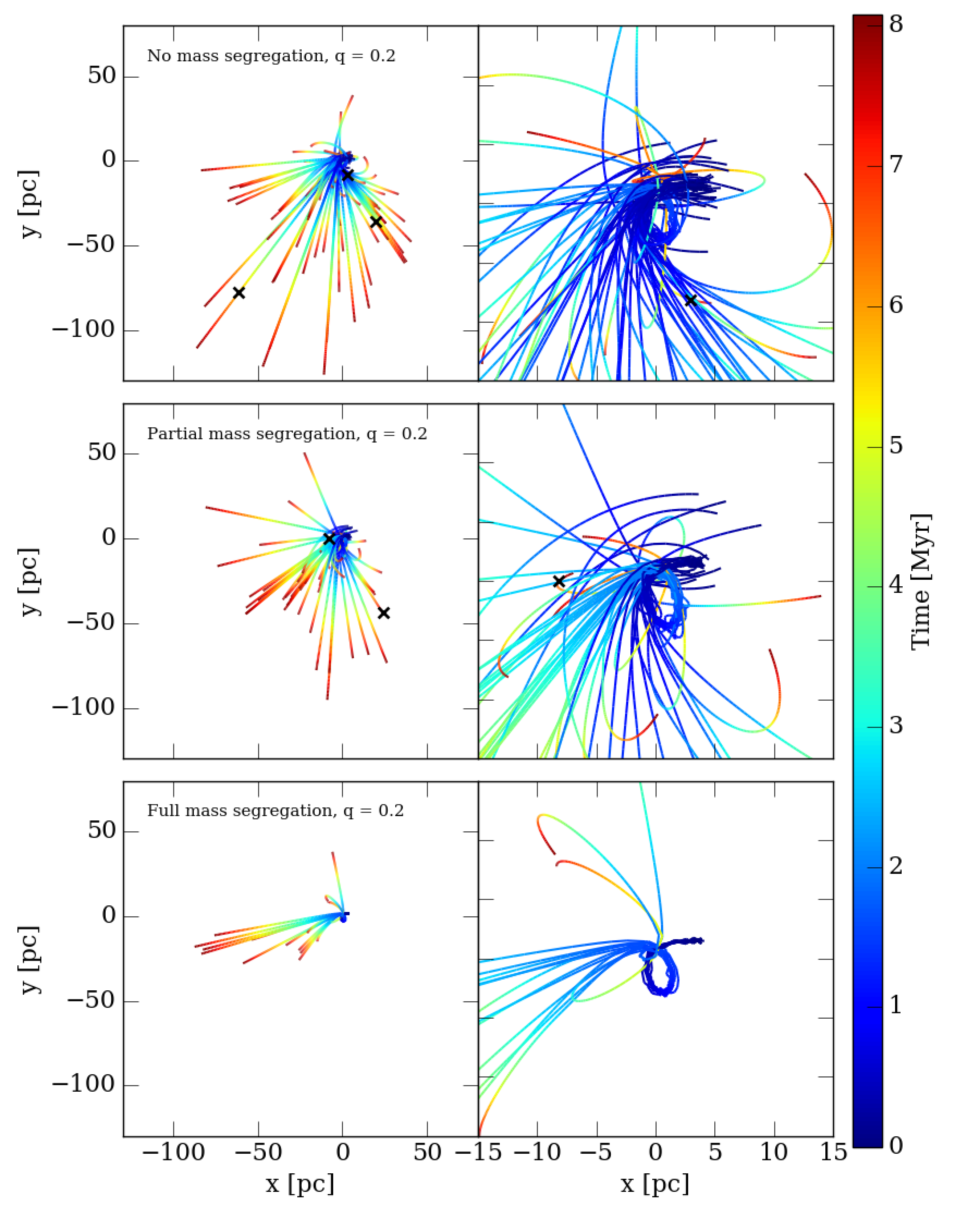}
\caption{Trajectories of massive stars ejected from fully-resolved tree-code cluster collision simulations. Shown are the trajectories of all stars originating in the secondary cluster with masses over $5M_\odot$ and final number densities below $1\,\mathrm{pc}^{-3}$. The right hand plots show the same data as those on the left, but are zoomed in on the region where the collision took place. A black cross marks the point where a star would have died in a supernova. When the clusters were not mass segregated, stars were less bound and many were ejected during the first pericentre passage. Some stars were far enough from the secondary cluster's core that they passed by the primary on the other side, giving them the opposite sense of rotation. When fully mass segregated, the massive stars were tightly held in the secondary cluster's core. Ejections were fewer and mostly took place during the second pericentre passage, leading to the directional distribution to be much more concentrated than in the other cases.}
\label{fig:trajectories}
\end{center}
\end{figure*}

The dispersal of the high-mass stars initially contained in the secondary cluster results in  a population of high-mass stars that appear to be isolated, i.e., they are no longer contained in dense stellar clusters. They are still in the general region of the rich stellar cluster that is produced in the merger process, but have been released into the field without an accompanying bound population of low-mass stars. 

Such a population is consistent with the observed ``isolated'' high-mass stars observed in the vicinity of 30 Doradus \citep{Bressertetal2012}.  Yet, their origin in this context can be understood to have been in a dense stellar cluster and their relative isolation is solely due to the subsequent cluster interaction,  tidal disruption and merger which released a number of the members of the secondary cluster into the field surrounding the resultant cluster.  Figure~\ref{fig:ninit_nfinal} compares the initial stellar densities of the higher mass ($m> 5 \solm$) stars from both the primary and secondary clusters with their final stellar density in the merged system. 
The stellar densities are calculated as the mean of the densities determined from the distances to the fifth, sixth and seventh nearest neighbour. 

From Figure~\ref{fig:ninit_nfinal}, we see that although the vast majority of the high-mass stars from both clusters have final stellar densities that are not too different from their initial values, there is a significant population of high-mass stars from the secondary cluster with final stellar densities 4-7 orders of magnitude lower than their initial values. These are the now isolated high-mass stars that have been dispersed into the field during the cluster merger. The results are very similar in form when the surface densities are calculated instead in 2-dimensional projection using the Cartesian axes as lines of sight. Only the value of the number density itself changes to account for increased crowding in projection, and even then the isolated stars are found at $\approx 0.1$ to $10\,\mathrm{pc}^{-2}$, clearly separated from the majority of the stars which remain roughly unchanged.

In the case of no or partial mass segregation, some stars from the primary cluster were also found by the simulation's end to have stellar densities below our $1\,\mathrm{pc}^{-3}$ threshold for isolation. These were however highly biased towards stars which had initial densities at the lower end of the range -- in other words, they were already at quite large distances from the core of the cluster, and so were easily shed. When fully segregated, isolated stars originated solely from the secondary cluster.

By the end of the simulations we found between $15$ and $119$ isolated high-mass stars, by our definition, in the set of nine simulations, as shown in Table~\ref{tab:sim_names}. These counts specifically exclude the stars which would already have gone supernova, though this was never a large number: the most removed in this way were six for the $q=0.1$ simulation with no mass segregation. In the fully segregated runs, no isolated stars at all had died by the final time, though the first isolated supernovae would be expected within 1 Myr of the end of the $q=0.3$ simulation. The main effect of the exclusion of dead stars was to lower the maximum mass of the isolated population, drastically for the non- and partially segregated runs where stars in the full range of the IMF were ejected, and to similarly lower the mean mass. Since lower mass stars were the bulk of the ejections in all cases, this second effect is not as severe.

The main source of variation in the number of isolated stars was the degree of mass segregation, and this will be discussed in Section~\ref{s:mass_seg}. An important note to make is that our clusters were set up using a concentration parameter $\epsilon = 1/3$, giving the primary and secondary equal densities at their half-mass radii. We noted in Section~\ref{ss:tidal_disruption} that there exists an anti-correlation in the cluster mass-radius relation. \citet*{GieMoeCla2012} also found that within $\sim 1\,\mathrm{Myr}$ clusters can reach a relation of approximately $r\sim M^{-1/3}$ via two-body relaxation, while \citet{GielesRenaud2016} found $r \sim M^{1/9}$ when combining two-body relaxation with repeated tidal shocking (though the time needed to reach this relation may be longer than the lifetimes of the O-stars). If we say that, realistically, a secondary cluster would be less dense than the primary, lowering the value of $\epsilon$, then the secondary cluster would be disrupted more easily than in our simulations. This means that the numbers of isolated high-mass stars we find are likely lower limits.

\subsection{Effects of mass segregation}\label{s:mass_seg}

The presence and degree of mass segregation in the clusters can have a significant effect on the ability of the encounter to disperse high-mass stars into the field. This may be understood in terms of how bound the high-mass stars are initially in the secondary cluster. Figure~\ref{fig:ninit_nfinal} shows that when the clusters were not mass segregated the number density $n_\mathrm{initial}$ initially ranged from $1$ to $10^6\,\mathrm{pc}^{-3}$ as they may be located anywhere in the cluster with equal probability. In contrast, in our fully mass segregated clusters, the high-mass stars are located only in the higher density cores with initial number densities in the range of $10^4$ and $10^5\,\mathrm{pc}^{-3}$. The maximum density is lower than before as the low-mass stars are no longer located in the core, and thus the total number of stars there is lower. The partially mass segregated clusters have their high-mass stars distributed with a similar range in $n_\mathrm{initial}$ as the non-segregated cluster but with a bias pushing the distribution to higher densities.

The main consequence of the mass segregation is that the high-mass stars are more likely to be contained deeper in the potential of the secondary cluster and hence require more energy from the encounter to unbind them. This decreases the number of high-mass stars that are dispersed into the field at each encounter such that the fully segregated clusters retain almost all their high-mass stars during the first closest approach. 

The nature of the cluster merger involves several closest approaches, with each subsequent one having a smaller closest approach and increasingly disrupting the secondary cluster.  This results in a tendency to disperse the stars in the core of the secondary cluster in the second or subsequent close passages. These episodic ejection events are directly linked to the cluster's inspiral and pericentre passages. This can be seen in Figure~\ref{fig:trajectories}: stars were quickly ejected from the unsegregated cluster while the stars in the fully segregated cluster remained in a coherent core. The majority of isolated stars were then ejected during the second pericentre, with a few more taking place after that. The ejections from the cluster core also gave these stars similar trajectories, while in the unsegregated case there was only a general preference for them to spread throughout the negative $y$ portion of the field. In particular, the massive stars on the cluster edge ended up moving in essentially every direction except that from which the secondary cluster approached the primary.

Table~\ref{tab:sim_names} shows the numbers and masses of stars in the isolated high-mass population at three times for each tree-code simulation. The numbers dropped consistently with increasing mass segregation. Averaging across the three mass ratios, at $8.08 \,\mathrm{Myr}$ there were $99$ isolated stars with no segregation, $59$ with partial segregation, and $15$ with full segregation.

The mean masses of the population are affected differently, falling by only a few solar masses as the degree of mass segregation increased. In general, the mean mass of an isolated high-mass star was $\approx 10 \solm$. The maximum mass found in each simulation followed yet another pattern, with the most massive isolated stars with no or partial mass segregation falling in a range from $20$ to $30 \solm$ depending on the mass ratio (though note that when the stars which had gone supernova by 8.08 Myr were included, the maximum masses fell in a range of $40$ to $90 \solm$ -- very massive stars were ejected, but they also died very early). With full mass segregation the range was only $9$ to $25 M_\odot$. Thus as long as a reasonable number of stars were unaffected by mass segregation it was possible for stars across the entire mass range to be liberated and later found as isolated stars. However, over time the number and maximum mass will fall as the highest mass stars begin to die.

While the fully mass segregated cluster produced very few isolated massive stars, it is unlikely that any real cluster resembles this case in which no massive stars at all are found outside the core and low-mass stars may only be found farther out. Rather, there is a preference for a central concentration of massive stars (\citealt{HilHar1998}; \citealt{Gouliermisetal2004}) as in our partially segregated case, and so we would expect instead several tens of isolated massive stars as predicted by these simulations.

\subsection{Kinematics of the dispersed high-mass population}

A constraint from the observations of \citet{Bressertetal2012} is that these stars should have fairly low velocities similar to those of stars in the originating cluster. OB runaways on the other hand, with velocities reaching beyond $100\,\mathrm{km}\,\mathrm{s}^{-1}$ (\citealt{Blaauw1961}; \citealt{CruzGonzalezetal1974}; \citealt*{Tetzlaffetal2011}), are common occurrence in dense stellar clusters due to interactions with high-mass tight binary systems that can eject stars at high-velocities from the cluster (\citealt{GiesBolton1986}; \citealt{FujiiZwart2011}). These OB runaways can be identified due to their high proper motions. In this paper, our fully-resolved tree-code gravity runs soften the gravitational forces within $10^4$ AU, ensuring that such high-velocity ejections do not occur. Instead, the increase in any given star's velocity is from the cluster encounter and merger.
 
Figure~\ref{fig:posvel} shows the velocity-position diagram of isolated high-mass stars in the non-mass segregated $q=0.2$ stellar system at 2, 4 and 8 Myr. As it was performed in the tree-code with $10^5$ particles and softened gravity ($10^4$ AU) there were no binary encounters or high-velocity runaways. The figure plots the magnitude of the stellar velocities of the isolated high-mass stars (meaning $n < 1\,\mathrm{pc^{-3}}$ and $m\ge 5 \solm$) as a function of the distance from the centre of the merged system. A significant number of stars have velocities of a few to nearly $20\,\mathrm{km}\,\mathrm{s}^{-1}$. Although not shown here, this is the same range of velocities possessed by the vast majority of massive stars remaining within the cluster, with a few stars reaching even higher values of nearly $50\,\mathrm{km}\,\mathrm{s}^{-1}$.

After 2 Myr some of the stars shown in Figure~\ref{fig:posvel} have reached distances between 20 and 30 pc; by the final time at $8\,\mathrm{Myr}$, they extend from 20 all the way to 140 pc away signifying that they have escaped the cluster. Their velocities represent the excess they were given on top of the escape velocity of the system, allowing them to reach these large distances.

It is of note that each of the three times shown in Figure~\ref{fig:posvel} the isolated stars may be grouped into two lines extending towards the upper right of the plot. These show that in this simulation the stars were cast off from the merging cluster in two distinct events. The stars' distances and velocities give approximately correct timings for the ejection events. It is thus possible to determine the number and timing of encounters within a cluster's recent history, as long as those encounters resulted in the dispersal of detectable stars. As an example, when the cluster was fully mass segregated, massive stars were only shed in the second encounter (as seen in Figure~\ref{fig:trajectories}) and so only the shorter line of stars is present.

\subsection{Supernova dispersal} \label{ss:sne}

\begin{figure*}
\begin{center}
\includegraphics[width=1.0\textwidth]{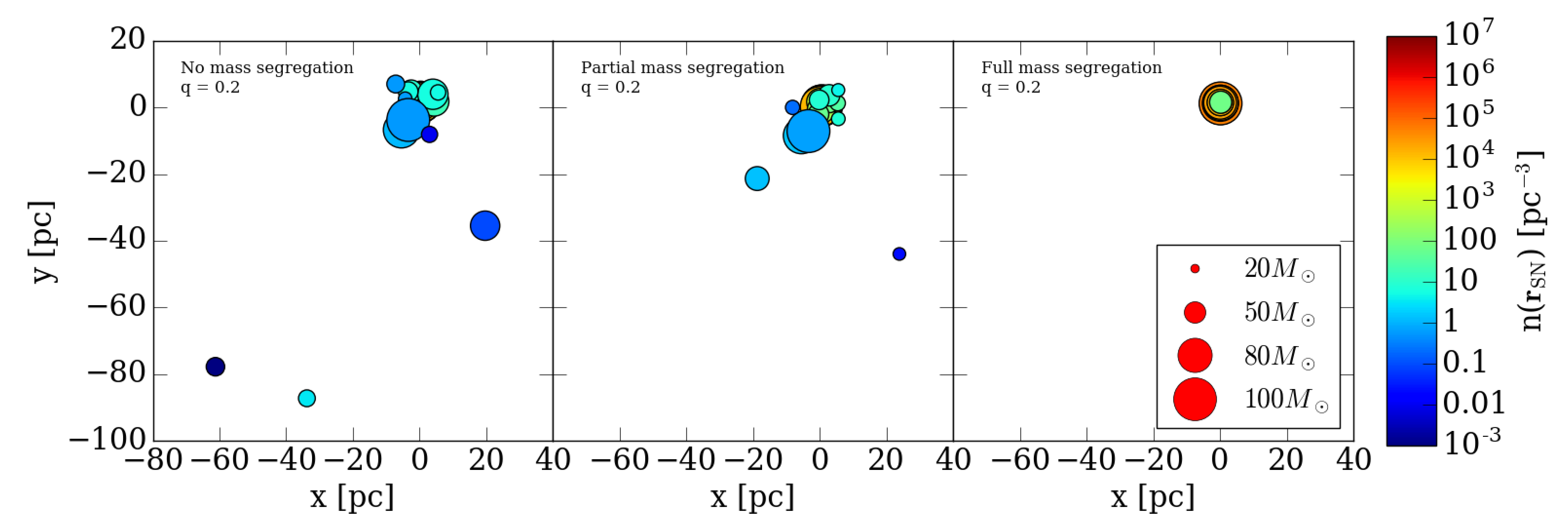}
\caption{The locations of supernovae (SNe) which detonated throughout the $8.08\,\mathrm{Myr}$ simulation runtime. The coordinates are the same as in Figures~\ref{fig:clustdisp} and \ref{fig:trajectories} with the origin at the primary cluster's initial centre of mass. The size of each point indicates the mass of the progenitor (and thus the time at which the SN occurred) while the colour indicates the number density $n$ of the star at the time of its death, with low $n$ events having been plotted on top. The majority of the supernovae occurred within or close to the clusters, but when mass segregation was not used or was only partial some massive stars could travel long distances and become isolated before they died. With full mass segregation, every SN took place within the merged cluster.}
\label{fig:sn_locations}
\end{center}
\end{figure*}

It has already been noted that over the $8\,\mathrm{Myr}$ timescale of the tree-code simulations we would expect high-mass stars to begin dying as supernovae (SNe). We have excluded these from our discussion so far, but it is of interest to see where and under what conditions the SN events would have taken place.

The particles representing the stars that would in reality be dying remained in place throughout the simulations, so our analysis here is entirely performed post-simulation. Setting the stellar ages to be equal to the simulation time (i.e. they were all formed at the beginning of the simulation), we then simply determined that a given star was dead for all simulation times greater than its lifetime as determined by its mass. We took the combined H- and He- burning timescales for stars initially rotating at $300\,\mathrm{km}\,\mathrm{s}^{-1}$ \citep{MeyMae2003} to be the lifetimes, interpolated for the stars in the simulations. With these values, stars of about $28 M_\odot$ or greater would be expected to die within the $8.08\,\mathrm{Myr}$ runtime.

Figure~\ref{fig:sn_locations} shows the locations of all $122$ supernovae taking place in the $q=0.2$ tree-code runs at the three levels of mass segregation. Furthermore we note the stellar number densities, calculated as before, in the environment of the supernovae at the time of the progenitors' deaths. Immediately noticeable is that in all three cases the vast majority of supernovae detonated within the merged cluster. When fully segregated, the result is predictably extreme, with the supernovae taking place solely in the core of the merged and the lowest number density being $75.5\,\mathrm{pc}^{-3}$, with the majority being on the order of $10^3$ or $10^4\,\mathrm{pc}^{-3}$.

While number densities for the supernovae in the core are still very high (and higher in some cases) in the runs with partial or no mass segregation, stars at higher masses were ejected in the merger and so have typically lower number densities. With partial mass segregation, three supernovae took place below $1 \,\mathrm{pc}^{-3}$ (one of which was $99.9 M_\odot$, visible in the plot just below the merged cluster), and with no mass segregation there were six. Some of the isolated supernovae occurred at distances of several tens of parsecs. These had lower mass progenitors, and so had more time in which to travel before they died.

It is interesting to note that the remnants of several supernovae are found throughout 30 Doradus which is $\approx 200 \,\mathrm{pc}$ in size, among them the remnants of SN1987A \citep{Townsleyetal2006}. Given that we see something not dissimilar to this in Figure~\ref{fig:sn_locations}, it may be that the cluster ejection mechanism is responsible for the observed distribution of remnants. Given the complexity of the region, with evolved stars in the field, several young clusters and ongoing star formation \citep{Sabbietal2016}, for now we merely note this as only a possibility.

\subsection{Binary stars}

A high fraction of massive stars are known to be members of multiple star systems -- the binary fraction of stars in 30 Doradus was measured by \citet{Sanaetal2013} to be $\approx 50$ per cent, though in other clusters this may be as high as 60 or 70 per cent \citet{SanEva2011}. Our simulations explicitly excluded binary stars to ensure that ejections would be solely due to the violent cluster merger. However, this has some implications for our final numbers of isolated stars, as a binary, even otherwise isolated, should not be counted.

The stars lost during the merger were those with lower specific orbital energies, so in the run without any mass segregation (i.e. without any correlation between stellar mass and orbital energy) the binary fraction of ejected stars would not be changed from that in the clusters. In contrast, mass segregation would lead to binaries being more tightly bound and we might expect instead that the binary fraction of ejected stars be lower than in the clusters.

At 8 Myr and assuming an upper limit on the binary fraction of $0.5$, the number of isolated massive stars should be at least $\approx 40$ to $60$ depending on the mass ratio $q$ without mass segregation, $\approx 20$ to $35$ with partial mass segregation, and $\approx 7$ to $8$ with full segregation. As above, with partial and full mass segregation the numbers may be higher if the segregation causes the preferential ejection of single stars.

\section{Comparison with ejections} \label{s:disp_ejec}

\begin{figure}
\begin{center}
\includegraphics[width=0.5\textwidth]{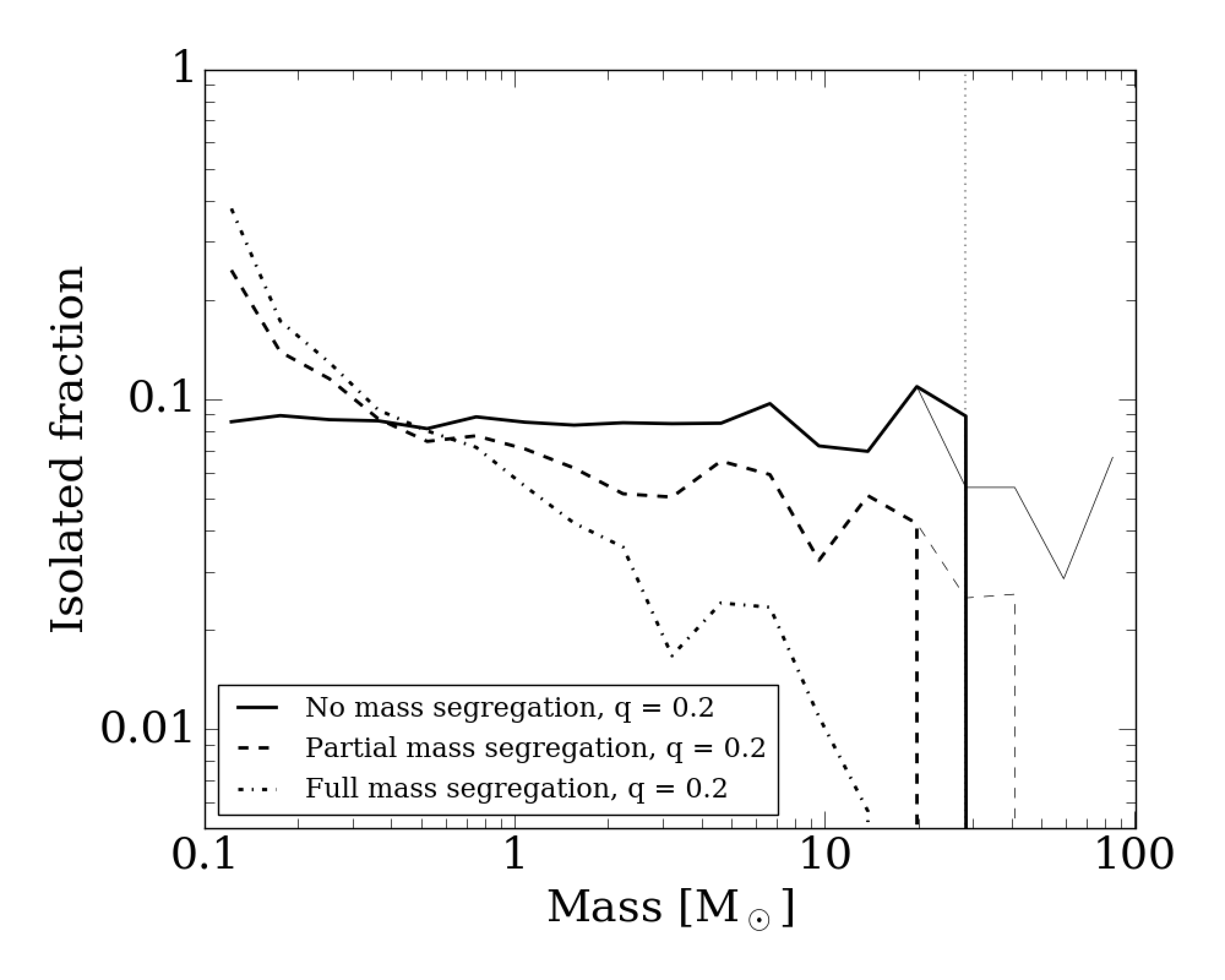}
\caption{The isolated fraction of stars as binned by their initial masses, plotted at the end time of 8.08 Myr. Stars belonging to both the primary and secondary cluster and over the full range of masses were included. The thicker lines show the isolated fraction of stars which would still be alive by this time, while the thinner lines also include the stars which would have already died; the vertical dotted line at $28.4 \,\mathrm{M}_\odot$ shows the most massive stars still alive.}
\label{fig:mass_dep}
\end{center}
\end{figure}

It is important to compare the results found with those of \citet{FujiiZwart2011} and \citet{BanKroOh2012} who reported on the production of isolated massive stars through binary ejections in young massive clusters. One feature of these authors' results is that stars were ejected without any directional preference. In contrast, our cluster merger simulations show strong preference for the stars to be dispersed in the tidal tails produced during the first and second pericentre passages -- see Figures~\ref{fig:clustdisp} and \ref{fig:trajectories}.

\citet{FujiiZwart2011} and \citet{BanKroOh2012} also noted that the isolated fraction of stars increased with stellar mass. This was simply calculated by binning the stars by their masses and then within each bin calculating the fraction determined to be isolated. Figure~\ref{fig:mass_dep} similarly shows the fraction for our $q = 0.2$ simulations with the three different levels of mass segregation, and also both including and excluding the stars which would have died by the time shown of 8.08 Myr.

The results reported here are very different from those of the previous authors. When there was no mass segregation, we found that isolated stars were produced equally effectively across the whole mass range at a fraction of about $0.085$, with growing noise at higher masses. When including the stars which would have died (above $28.4\,\mathrm{M}_\odot$) a small downturn can be seen.

The simulations in which mass segregation was implemented whether partially or fully display a negative trend with mass, strongly contrasting with the results of \citet{FujiiZwart2011} and \citet{BanKroOh2012}. The isolated fraction of stars in the lowest mass bins in the simulations with both partial and full mass segregation were $0.246$ and $0.378$ respectively. The fraction fell rapidly with increasing mass and reached zero within a few tens of solar masses. The slope was steepest in the fully segregated run. This is a natural consequence of the segregation: when in effect, massive stars were the most bound and so harder to eject during the cluster merger, while the inverse was true for low mass stars.

Figure~\ref{fig:mass_dep} only shows the results for $q=0.2$, but they are similar for both $q = 0.1$ and $0.3$. Generally the lines for each fraction are shifted up and down by $\approx 10^{-2}$ in each case. This can be explained in the same manner as the numbers of ejected stars in Table~\ref{tab:sim_names} -- the majority of isolated stars originated from the more easily disrupted secondary cluster, so as the number of stars in the cluster increased more were available to be dispersed.

Since we would expect a realistic cluster to be at least somewhat mass segregated, these results indicate that a pure dispersal scenario would be very distinct from a binary ejection scenario, with coherent spatial distributions of isolated massive stars being produced alongside a large number of lower mass stars, while binary ejections produce stars moving in all directions and favour high mass stars. However, since for these simulations we intentionally removed close interactions, we do not observe the binary ejections that would likely occur alongside a real cluster merger and the subsequent dispersal of massive stars.

\section{Conclusions}

The nature of cluster formation whereby larger clusters grow from the merger of smaller systems is a generally ignored aspect of star and cluster formation. In the later stages of the process, the mergers are liable to be dry as the natal gas has either been accreted or dispersed from the system. In such cases, the merger process needs to release some of the kinetic energy of the system, in the form of tidal tails or the tidal disruption of the lower-mass cluster. We have modelled this process through direct (lower resolution) $N$-body and full resolution tree-code simulations of the cluster merger process. In particular, the tree-code simulations used softened gravity within $10^4\,\mathrm{AU}$, ensuring that we would only find isolated stars from the merger process and not from binary ejections. The merger process results in a population of the stars escaping the system with the excess kinetic energy. These stars are generally given a a velocity boost of order $10\,\mathrm{km}\,\mathrm{s}^{-1}$, corresponding to the cluster interaction at closest approach, and form an envelope of slowly escaping stars.

The escaping population contains a significant number of high-mass stars ($m\ge 5 \solm $). These high-mass stars can come from any part of the cluster, as their ejection from the system is due to the tidal disruption of the secondary cluster. Their velocities of $5-20\,\mathrm{km}\,\mathrm{s}^{-1}$ allow them to reach distances of up to 140 pc within 8 Myr of the interaction. They attain very low stellar densities of between 1 and $10^{-3}$ stars pc$^{-3}$, making them isolated high-mass stars, in contrast to their clustered origin. Their low velocities ensure they are not classified as high-mass runaways that arise from binary ejections. We propose that this is a potential origin of the seemingly isolated massive stars reported in 30 Doradus by \citet{Bressertetal2012}.

The degree of mass segregation in the merging clusters is an important factor. As a greater fraction of massive stars are concentrated in the clusters' cores, fewer of them will be dispersed into the field. In perhaps the most realistic case for mass segregation, with half the stars segregated, $32$, $60$ and $71$ massive stars that had not yet died as supernovae were isolated by $4\,\mathrm{Myr}$ for mass ratios $q=0.1$, $0.2$ and $0.3$ respectively. In turn these convert to $25.8\%$, $24.5\%$ and $22.3\%$ of the massive stars in the secondary cluster being dispersed, though these numbers are specific to our initial conditions. These are conservative values as we set our clusters to have equal densities at their half-mass radii, while in reality the secondary cluster could be less dense and so easier to disrupt.

The merger events leave kinematic signatures in the escaping stars as they all are given the excess kinetic energy at the same point in time, and thus have distances from the cluster directly proportional to their escaping velocities. The (dry) merger histories of clusters can thus be reconstructed through the properties of this population of slowly escaping stars around the cluster. The massive stars are furthermore only the high-mass component of one or more tidal tails stretching all the way back to the original clusters, by now probably fully merged. These signatures stand in contrast to the random directions of the ejected stars produced in binary scatterings found by \citet{FujiiZwart2011} and \citet{BanKroOh2012}, another avenue for the production of isolated massive stars. These authors found that high-mass stars would be preferentially ejected over low-mass stars; our simulations found that, conversely, stars were dispersed independently of mass, with the only factor being how bound each star is to the merging cluster. The HST images of \citet{Sabbietal2012} do resemble the tails formed in our simulations. The HST astrometry of \citet{Plataisetal2015} and future kinematic information made available by Gaia DR2, combined with line-of-sight velocities, should allow discrimination between the binary ejection and cluster dispersal scenarios.


\section*{Acknowledgements} 

WEL and IAB gratefully acknowledge support from the ECOGAL project, grant agreement 291227, funded by the European Research Council under ERC-2011-ADG. MR acknowledges funding from the Nadacia SPP grant No. 28/2013. MG acknowledges support from the Royal Society in the form of a University Research Fellowship (URF) and the European Research Council (ERC-StG-335936, CLUSTERS). This work used the compute resources of the St Andrews MHD Cluster. The authors thank the anonymous referee for their suggestions which have greatly improved this paper's quality.


\bibliography{iab}

\bsp

\label{lastpage}

\end{document}